\documentclass[
prl,
reprint,
showpacs,
amsmath,amssymb,
aps,
floatfix,
]{revtex4-1}

\usepackage{graphicx}
\usepackage{dcolumn}
\usepackage{bm}
\usepackage{subfigure}

\newcommand{\bol}[1]{\boldsymbol{\mathrm{#1}}}
\newcommand{\derivative}[2]{\frac{{\rm d} #1}{{\rm d} #2}}
\begin{document}

\title{
Multiscale Simulation of
History Dependent Flow in Polymer Melt}
\author{Takahiro Murashima}
\email{murasima@cheme.kyoto-u.ac.jp}
\author{Takashi Taniguchi}
\email{taniguch@cheme.kyoto-u.ac.jp}
\affiliation{
Department of Chemical Engineering, Kyoto University, Kyoto 615-8510,%
Japan}
\affiliation{
CREST, JST, Kawaguchi, Saitama, 332-0012, Japan%
}
\date{\today}

\begin{abstract}
We have developed a new multiscale simulation technique
to investigate history-dependent flow behavior of entangled polymer melt,
using a smoothed particle hydrodynamics simulation with
microscopic simulators 
that account for the dynamics of entangled polymers
acting on each fluid element.
The multiscale simulation technique is applied to 
entangled polymer melt flow around
a circular obstacle in a two-dimensional periodic system.
It is found that
the strain-rate history-dependent stress of the entangled polymer melt
affects its flow behavior,
and the memory in the stress causes 
nonlinear behavior even in the regions where ${\rm Wi} \le 1$.
The spatial distribution of the entanglements $\langle Z \rangle$ is 
also investigated. 
The slightly low entanglement region is
observed around the obstacle and is found to be broaden 
in the downstream region. 
\end{abstract}

\pacs{83.10.Rs,83.10.Mj,83.60.Df,83.80.Sg}
\maketitle

Industrial products using polymeric materials 
have become increasingly integral to our lives.
One of the important characteristics of
thermo-plastic polymeric material 
is that it can be easily molded and processed 
by controlling its state from a solid to a melt,
which is beneficial in
a variety of practical applications.
The melt state of polymeric material can exhibit
a lot of characteristic phenomena, 
e.g.,
die-swell and rod-climbing\cite{Boger_Walters},
depending on the dynamic response of
the polymer's microscopic internal states
under an imposed strain or strain-rate. 

Predicting the flow of polymeric fluid 
is difficult
because the macroscopic flow behavior 
depends heavily on the dynamic behavior of
the microscopic internal states,
and the states of polymers 
with very high molecular weight 
are complex at the microscopic level.
When a polymer melt consists of polymers 
with a molecular weight $M$ higher than 
the entanglement molecular weight $M_{\rm e}$,
the polymer molecules are entangled with each other,
and
the relaxation time of the polymer conformation is long
compared to that of dilute polymer solutions 
because of the entanglement.
Therefore,
it is difficult
to predict the rheological behavior, 
even for a homogeneous bulk entangled polymer melt,
using molecular dynamics approaches
without any coarse-graining procedures for
the microscopic internal degrees of freedom.
Reptation theory\cite{Edwards1967,deGennes1971,DE1986}, 
in which a polymer chain is coarse-grained 
to a tube confined by the surrounding polymer chains,
explains the complex behavior of 
entangled polymer melt composed of mono-disperse linear polymers.
To apply the concept of a tube
to a wider variety of polymers,
the original reptation theory has been
improved by
introducing several important physical 
mechanisms\cite{MLD1998,IM2001,IM2002,Likhtman2003}.
Extended reptation theory
based on the Fokker-Planck equation 
for tube segments
succeeds
in explaining 
many experimental results\cite{MLD1998,IM2001,IM2002,Likhtman2003}.
However, 
it is difficult to apply the theory
to arbitrary polymer melts
because
it is too mathematically complex to incorporate
the molecular pictures of polymers 
of arbitrary architectures
into the Fokker-Planck equation. 
On the other hand,
Langevin stochastic models based on reptation theory
using the complex molecular configurations
have been developed 
and
can reproduce the rheological properties of 
various polymeric materials
with branches and/or molecular weight 
distributions\cite{Masubuchi2001,SLTM2001,DT2003,Likhtman2005}.
These models
use efficient numerical computation
to predict the bulk rheology of 
polymer melts; 
however, they are not applicable to 
macroscopically inhomogeneous flows
only by themselves. 
The macroscopic flow behavior of polymer melt is usually predicted 
using a fluid dynamics simulation with
a constitutive equation that describes a nonlinear 
and history dependent relationship
between the stress and strain-rate fields 
to represent complex fluids without microscopic details.
However, no general constitutive equation that is applicable 
to entangled polymer melts with arbitrary polymer architecture
exists because a general polymer melt, 
e.g., polymers with branches and/or molecular weight distributions,
has extremely complex molecular states,
which makes the stress response unpredictable
even in simple flow patterns.
As described above,
each of the microscopic and macroscopic approaches has 
limited applications.
Recently, a molecularly derived constitutive 
equation technique
is developed for low-molecular polymer melts 
without entanglements\cite{Ilg2011}.
If it is generalized to a variety of polymer architectures,
the technique can be attractive and useful.
However, it is in an early phase of development toward generalization.  

As an alternative to using a constitutive equation,
we propose a new simulation technique that incorporates Langevin 
course-grained simulators\cite{Masubuchi2001,SLTM2001,DT2003,Likhtman2005} 
into a fluid dynamics simulation.
This approach is a type of micro-macro simulations that 
were mainly developed
for dilute polymer solutions in which the stress tensor 
of each fluid element is obtained
from a microscopic simulation\cite{CONNFFESSIT,LPM,BCF},
and applied not to dilute polymer solutions 
but to well entangled polymer melts
here.
In contrast to the dilute polymer solutions, 
the polymer melts have entanglement interactions 
between neighboring polymers.
In order to investigate the entangled polymer melts using
the micro-macro simulation,
we need to satisfy 
(i) a local stress tensor is represented by
an ensemble of a sufficiently large number 
of entangled polymers
and to consider (ii) strong correlation among the polymers in an ensemble
because of the entanglements.
The advection of the stress, $(\bol{v}\cdot \bol{\nabla})\bol{\sigma}$,
in the Eulerian description is well established in any fields.
However,
it only advects the macroscopic quantity of the stress 
(the statistical conformation tensor)
without involving microscopic molecules themselves.
Because the correlations among polymers,
namely the entanglements,
intrinsically affect their macroscopic properties 
in the entangled polymer melts, 
the advection of an ensemble of polymers correlating 
at the microscopic level should be consistently managed
with the advection of the macroscopic variables.
Therefore, we adopt a Lagrangian fluid particle 
method\cite{EKH2002,EEF2003,MT2010}
to ensure the advection of the ensemble of entangled polymers.
Each ensemble of entangled polymers
is assigned on each fluid element.
The advection of microscopic details
is essential to the description of the macroscopic flow behavior
for entangled polymer melts 
because the conformations of entangled polymers 
are constrained by and correlated to the surrounding polymers.
Note that 
the Eulerian techniques 
can still be useful even for entangled polymer melts 
when the system has a translational symmetry\cite{YY2009,YY2010}.

To maintain the information pertaining to entanglement 
and deformation in polymers,
we perform a Lagrangian fluid particle simulation 
in which each fluid particle has 
a microscopic level simulator 
that accounts for the internal states of the fluid particle\cite{MT2010}.
Assuming that 
the polymer melt is an isothermal and incompressible fluid,
the governing equations 
for the $i$-th fluid particle that constitutes
the polymer melt are given
by the following equations;
\begin{align}
\rho_i 
\derivative{\bol{v}_i}{t}
=\bol{\nabla}\cdot\bol{\sigma}_i-\bol{\nabla}p_i
+ \bol{F}^{\rm b},&\quad
\derivative{\bol{r}_i}{t}
=\bol{v}_i,\label{vel_dt}\\
\bol{\sigma}_i\equiv\bol{\sigma}_i(\mathcal{Q}_i),&
\label{stress}
\end{align}
where $\bol{F}^{\rm b}$ is a body force
and the pressure $p_i\equiv p_i(\{\rho_i\})$ is properly considered
to guarantee the incompressibility.
These equations are solved using macroscopic variables,
except for Eq. (\ref{stress}). 
The local stress tensor $\bol{\sigma}_{i}$ 
depends on a microscopic ensemble $\mathcal{Q}_i$ of entangled polymers,
which represent entangled states of polymers
under 
an instantaneous
local strain-rate $\bol{\kappa}_i \equiv (\bol{\nabla}\bol{v})^{\rm T}_i$.

The slip-link model\cite{SLTM2001,DT2003} is 
a simulation model 
that can accurately describe the dynamics of entangled polymers.
The model is composed of confining tubes with some
entanglement points, called slip-links, which confine a pair of polymers
and represent 
effective constraints in virtual space. 
The average number of slip-links or entanglements 
on a polymer at the equilibrium state is represented as 
$\langle Z \rangle_{\rm eq}\equiv M/M_{\rm e}$.
In the simulation, we trace the configurations of confining tubes
constrained by the slip-links.
The slip-links are relatively convected each other
and 
the confining tubes are deformed according 
to the macroscopically obtained 
local velocity gradient tensor $\bol{\kappa}$.
The reptations of polymers 
generate or eliminate slip-links. 
For given chain configurations,
the stress tensor $\bol{\sigma}^{\rm p}$ 
derived from deformations of entangled polymers
is calculated from
$\sigma_{\alpha \beta}^{\rm p} =\sigma_{\rm e} \sum_j \langle
r_{j\alpha}^{\rm s}r_{j\beta}^{\rm s}/|\bol{r}_j^{\rm s}|\rangle/a_{\rm s}$ 
where $a_{\rm s}$ is the unit length of the slip-link model and
$r_{j\alpha}^{\rm s}$ is the
$\alpha$-component of the $j$-th tube segment vector connecting
adjacent slip-links along a polymer. 
The unit of stress $\sigma_{\rm e}$ in the slip-link model relates 
to the plateau modulus $G_{\rm N}$ 
as follows: $\sigma_{\rm e}=(15/4)G_{\rm N}$\cite{DT2003}.
The slip-link model has two characteristic time-scales: 
the Rouse relaxation time
$\tau_{\rm R}$ and the longest relaxation time $\tau_{\rm d}$.
The Rouse relaxation time $\tau_{\rm R}$
and the longest relaxation time $\tau_{\rm d}$ 
relate to $\langle Z\rangle_{\rm eq}$ as follows:
$\tau_{\rm R}=\langle Z\rangle_{\rm eq}^2 t_{\rm e}$ 
and $\tau_{\rm d}\propto 
\langle Z\rangle_{\rm eq}^{3.4}t_{\rm e}$\cite{DT2003,DE1986},
where $t_{\rm e}$ is the time unit of the slip-link model.
The contour length relaxation of a confining tube occurs 
on the time-scale of $\tau_{\rm R}$,
while the orientational relaxation occurs on the time-scale of $\tau_{\rm d}$.
These two characteristic times appear
in the stress relaxation.

Each polymer simulator describing a fluid particle 
computes the polymer configurations
at each time step, 
and the recorded configurations are used as the initial conditions 
of the next time step.
Typically, the macroscopic time unit $t_{\rm macro}$
and microscopic time unit $t_{\rm micro}$
have a large time-scale gap, and therefore
the macroscopic time unit $t_{\rm macro}$ must be divided
into $N t_{\rm micro}$.
Because the slip-link model used here is sufficiently coarse-grained 
and the time unit $t_{\rm e}$ 
can be the same as the time-scale of the macroscopic fluid
$t_{\rm macro}$,
we employ 
$t_{\rm macro}=t_{\rm micro}\equiv t_{\rm e}$.

Note that we set the local stress of the macroscopic fluid to
$\bol{\sigma}=\bol{\sigma}^{\rm p}+\bol{\sigma}^{\rm d}$,
where $\bol{\sigma}^{\rm d}$ is an extra dissipative stress tensor.
Because the slip-link model is a Langevin coarse-grained model 
based on reptation theory,
microscopic dynamics smaller than a tube segment are treated as 
a random force exerted on a slip-link,
and 
the contribution from the microscopic dynamics 
does not explicitly appear in the stress tensor of the slip-link
model.
We assume the dissipative stress $\bol{\sigma}^{\rm d}$ 
to be the Newtonian viscosity $\eta_{\rm_d}\bol{D}$,
where 
$\bol{D}\equiv \bol{\kappa}+\bol{\kappa}^{\rm T}$ 
is a strain-rate tensor.

The main procedures of our simulation are summarized as follows:
(1) Update $\{\bol{v}_i\}$, $\{\bol{r}_i\}$ at the macroscopic level.
(2) Calculate $\{\rho_i\}, \{\bol{\kappa}_i\}$ at the macroscopic level.
(3) Obtain $\{p_i\}$ from the density distribution $\{\rho_i\}$
at the macroscopic level.
(4) Update the local slip-links of the $i$-th fluid particle
under the local strain-rate
$\bol{\kappa}_i$ 
and then obtain $\bol{\sigma}_i$ from the resulting configuration 
of the slip-link model.
This procedure is executed on each fluid particle in turn.
(5) Calculate $\{\bol{\nabla}\cdot\bol{\sigma}_i\}$ 
      and $\{\bol{\nabla}p_i\}$ at the macroscopic level.
(6) Return to (1).

We update 
$\{\bol{v}_i\}$, $\{\bol{r}_i\}$ by integrating Eq.~(\ref{vel_dt}).
We calculate the density 
at the position of
each particle in the new configuration
using a method in
the usual smoothed particle hydrodynamics technique\cite{SPH}:
$\rho_i = \sum_j m_j W(|\bol{r}_j-\bol{r}_i|,h)$,
where $m_i$ is the mass of the $i$-th particle 
and $W(|\bol{r}|,h)$ is a Gaussian-shaped function with cutoff length $2h$.
The deviation of the local density from the initial constant density 
$\rho_0$ 
results in a local pressure force $-\bol{\nabla}p$.
To obtain the spatial derivative of the velocity field, stress field, and
pressure field ($\bol{\nabla}\bol{v}, \bol{\nabla}\cdot\bol{\sigma}, 
\bol{\nabla}p$), 
we use a technique that was developed
for modified smoothed particle hydrodynamics\cite{MSPH,FPM}.

\begin{figure}
\includegraphics[width=1.\columnwidth]{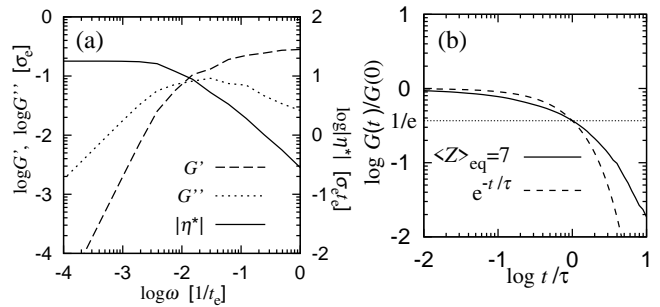}
\caption{\label{fig:g1g2}
(a) Rheological data and 
(b) relaxation modulus $G(t)$ 
obtained from the slip-link simulation
with $\langle Z\rangle_{\rm eq}=7$ in the bulk.
The storage modulus $G'$ (dashed line), 
the loss modulus $G''$ (dotted line), 
and the complex viscosity $\eta^*$ (solid line)
are plotted against the angular frequency $\omega$ in (a).
The magnitude of the complex viscosity, $|\eta^*|=\sqrt{G'^2+G''^2}/\omega$,
is considered to be the shear viscosity $\eta$ 
of the low shear-rate region
(Cox-Merz rule).
The normalized relaxation modulus $G(t)/G(0)$ (solid line) 
and
a single exponential curve with a relaxation time $\tau$ 
(dashed line) are plotted against the normalized time $t/\tau$
 in (b).
}
\end{figure}

To demonstrate the efficiency of the proposed multiscale simulation,
we consider a system in which the flow history 
can affect the flow behavior.
One such system is 
a polymer melt flow around
an infinitely long
cylinder oriented in the $z$-direction with a
radius $r_{\rm c}$,
which
flows in the $x$-direction. 
Because of the symmetry of the system,
we can treat the system as two-dimensional,
and the flow can be described as two-dimensional flow 
in the $xy$-plane.
We assume 
a non-slip boundary condition 
for the velocity on the surface of the cylinder
and periodic boundary conditions at the boundaries of the system.
The dimensionless parameters governing the problem are 
the Reynolds number
${\rm Re}=\rho |\bar{\bol{v}}| r_{\rm c} / \eta^0$,
the Weissenberg number ${\rm Wi}=\tau_{\rm d} D_{xy}$,
and the viscosity ratio $\eta_{\rm d}/\eta^0$,
where $\bar{\bol{v}}$ is the average flow velocity. 
The zero shear viscosity $\eta^0$ of the polymer melt is 
given by $\eta_{\rm p}^0+\eta_{\rm d}$, 
where
 $\eta_{\rm p}^0$ is the zero shear viscosity 
of a polymer melt described only by the slip-link model.
From the rheological data shown in Fig.~\ref{fig:g1g2} (a),
which were derived from the slip-link simulation 
with $\langle Z\rangle_{\rm eq}=7$ in the bulk, 
we obtain
the longest relaxation time $\tau_{\rm d}\simeq 200 t_{\rm e}$ 
and the zero shear viscosity 
$\eta_{\rm p}^0 \simeq 17.5 \sigma_{\rm e}t_{\rm e}$. 
The cylinder radius $r_{\rm c}$ was set to $3a$,
where $a$ is the unit length in the fluid particle simulation,
and we assign the unit mass $m$ to all fluid particles.
The wall of the cylinder
consists of fixed fluid particles evenly spaced 
on the perimeter.
Each fluid particle consists of 10000 polymers, 
enough to 
describe the bulk rheological properties of the polymer melt
under an imposed shear and/or extensional deformation\cite{MT2010}.
About 900 fluid particles with 10000 polymers each
are evenly placed initially in the system 
and then move according to Eq. \eqref{vel_dt}; therefore,
we need to simultaneously solve 
the dynamics of 9000000 polymers.
Because the diffusive motion of the center of mass of a single polymer
is negligible compared to the translational motion of 
the center of mass of an ensemble of entangled polymers,
we neglect any transportation of polymers 
between adjacent fluid particles. 
With this assumption,
each slip-link simulation can be performed independently of the others,
making parallel computing effective
in this multiscale simulation.

Under the body force 
$\bol{F}^{\rm b}/(\eta^0 /a t_{\rm e})=(5.0\times 10^{-4},0)$,
the flow 
becomes steady-state in about $1000 t_{\rm e}$.
The average flow velocity $\bar{\bol{v}}$ in steady-state in this system
is nearly equal to $(0.04, 0)a/t_{\rm e}$ for a fully Newtonian flow
$\eta^0=\eta_{\rm d}$ ($\bol{\sigma}^{\rm p}=0$), and $(0.055, 0)a/t_{\rm e}$ 
for a polymer melt flow with $\langle Z\rangle_{\rm eq}=7$ 
and $\eta_{\rm d}/\eta^0=0.1$.
In both cases,
${\rm Re}$ is less than 0.2, and the flow is laminar. 
In the polymer melt case, 
the average flow velocity is higher than that of Newtonian flow,
i.e.,
the flow exhibits shear thinning behavior
because of ${\rm Wi} > 1$ in the vicinity of the cylinder.

\begin{figure}
\includegraphics[width=1.\columnwidth]{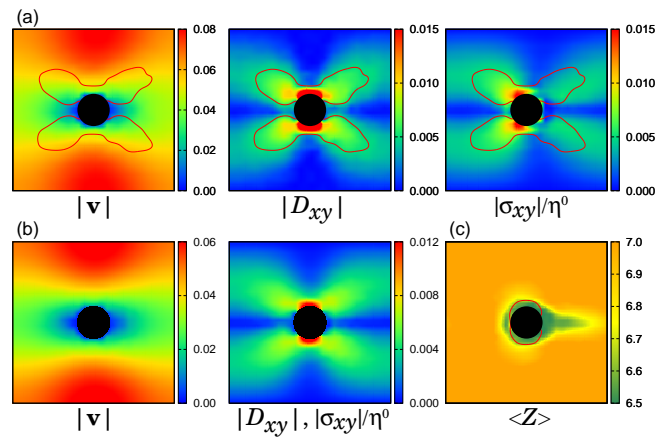}
\caption{\label{fig:contour_map}Color contour maps of time-averaged 
magnitudes of the velocity field $|\bol{v}| [a/t_{\rm e}]$,
strain-rate $|D_{xy}| [1/t_{\rm e}]$, 
and shear stress over the zero shear viscosity
$|\sigma_{xy}|/\eta^0 [1/t_{\rm e}]$
for (a) the polymer melt with $\langle Z\rangle_{\rm eq}=7$ 
and $\eta_{\rm d}/\eta^0=0.1$ 
and 
(b) the Newtonian fluid with $\eta^0=\eta_{\rm d}$
in steady-state.
Figure (c) shows $\langle Z \rangle$ in the polymer melt.
The regions inside the red lines in (a)
correspond to ${\rm Wi} \ge 1$ and those in (c) to $\tau_{\rm
 R}\dot{\gamma} \ge 1$ where $\dot{\gamma} \equiv\sqrt{{\rm Tr} (\bol{\kappa}^{\rm T}\bol{\kappa})}$.
}
\end{figure}

To investigate the velocity field $\bol{v}$, 
strain-rate $D_{xy}$, and shear stress $\sigma_{xy}$
in steady-state,
we employ a linear interpolation to transform the data 
at the particle positions into the values at regular lattice points
and then time-average
the data evaluated at the lattice points.
To eliminate the noise of the data, the time-averaging was
carried out from 1000$t_{\rm e}$ to 2000$t_{\rm e}$.
Figure~\ref{fig:contour_map} shows 
the spatial distributions of 
$|\bol{v}|$, $|D_{xy}|$ and $|\sigma_{xy}|/\eta^0$ in steady-state for
(a) the polymer melt with $\langle Z\rangle_{\rm eq}=7$ 
and $\eta_{\rm d}/\eta^0=0.1$
and (b) the Newtonian fluid with $\eta^0=\eta_{\rm d}$.

Reflecting laminar behavior,
the magnitudes of $\bol{v}$ 
and $D_{xy}$ in Fig.~\ref{fig:contour_map}
appear to be nearly symmetric
between the upstream and downstream regions
unlike $\sigma_{xy}$. 
In Fig.~\ref{fig:contour_map} (a), 
the nonlinear relationship
between $\sigma_{xy}$ and $D_{xy}$
is observed near the cylinder where ${\rm Wi}>1$
because of shear thinning.
Moreover, the shear stress $\sigma_{xy}$ of the polymer melt 
exhibits an apparent asymmetry
between the upstream and downstream regions.
In general,
the viscoelastic stress relates to the shear-rate
through the relaxation modulus
according to the following equation:
$\sigma(t)= \int_{-\infty}^{t} dt' G(t-t') \dot{\gamma}(t')$;
the stress depends on the flow history.
\begin{figure}
\includegraphics[width=1.0\columnwidth]{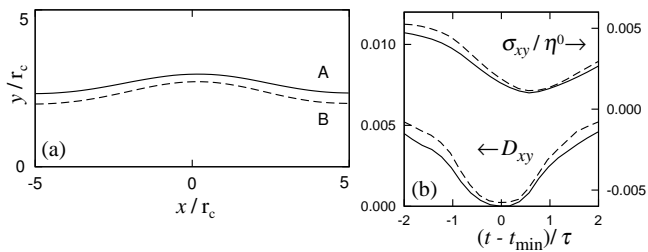}
\caption{\label{fig:ryuusen_data}
Stream lines 
(solid line(A) and dashed line(B)) obtained 
from the time-averaged 
velocity field with particle tracking are plotted 
against position in (a).
These stream lines are in the region where ${\rm Wi} \leq 1$.
The strain-rate $D_{xy}$ 
and the shear stress $\sigma_{xy}$
along the stream lines
are
plotted against
the elapsed particle tracking time in (b).
Each line type corresponds to each stream line.
}
\end{figure}
We investigate the behavior of $\sigma_{xy}$ and $D_{xy}$
on two typical stream lines 
(A: solid line and B: broken line 
shown in Fig.~\ref{fig:ryuusen_data} (a))
in the region where ${\rm Wi} \leq 1$.
Figure ~\ref{fig:ryuusen_data} (b)
shows $\sigma_{xy}$ and $D_{xy}$
along these stream lines and
plots them against the elapsed particle tracking time.
The time origin $t=0$ is set to $t_{\rm min}$, 
when the strain-rate $D_{xy}$ is at a minimum in each stream line.
A nonlinear relationship 
between $\sigma_{xy}$ and $D_{xy}$ 
is not 
evident in either stream line
because ${\rm Wi} \leq 1$;
however, the minimum of $\sigma_{xy}$ 
appears to be shifted from that of $D_{xy}$ 
by a time difference $t_{\Delta}$.
Applying a step shear strain $\gamma=0.5$ to the slip-link model
with $\langle Z\rangle_{\rm eq}=7$ in the bulk,
we obtain the relaxation modulus $G(t)\equiv\sigma(t)/\gamma$ 
shown in Fig.~\ref{fig:g1g2} (b).
The stress relaxation time $\tau$ is estimated to be
$\tau \simeq 50 t_{\rm e} (\simeq \tau_{\rm R})$
when $G(t=\tau)/G(0)=1/e$.
The time difference $t_{\Delta}$ is found to be nearly equal to $\tau$.

Finally,
we investigate
the spatial distribution of entanglements $\langle Z \rangle$
shown in Fig.~\ref{fig:contour_map} (c).
In the vicinity of the cylinder, the entanglements slightly decrease.
When $\tau_{\rm R}\dot{\gamma} \ge 1$,
polymer's contour length is highly extended. 
The extended polymer longer than $\langle Z\rangle_{\rm eq} a$
is easy to shrink, which causes disentanglement.
In the upstream and downstream regions around the cylinder,
the reduction of $\langle Z \rangle$ is also observed 
because of nonzero $|D_{yy}|$. 
The flow advection broadens the region where 
$\langle Z \rangle < \langle Z \rangle_{\rm eq}$ 
in the downstream region.
The tail length in the downstream region is roughly estimated
to be $\tau_{\rm R}|\bar{\bol{v}}| \sim r_{\rm c}$.

Using the Lagrangian particle method 
to trace and maintain the entire configurations of polymers,
we have been able to describe the memory effect
in the polymer melt flow.
The presented multiscale simulation 
is applicable to various polymer melts,
because the slip-link model 
or the alternative course-grained models\cite{Masubuchi2001,Likhtman2005}
can address a variety of polymer architectures,
e.g., linear and/or branched polymers, polymer blends, and
polydispersed polymers.
The multiscale simulation 
is advantageous because it employs 
a fully Lagrangian method at the macroscopic level,
while
conventional micro-macro techniques 
which have difficulties 
accounting for the macroscopic advection of
microscopic internal states.

\begin{acknowledgments}
We thank Professor R. Yamamoto and Dr. S. Yasuda 
for their diligent discussions and helpful advice.
We greatly appreciate Professor M. Doi 
and
Professor J. Takimoto 
providing their source
code of the slip-link model included 
in the {\it OCTA} system ({\it http://octa.jp/}).
\end{acknowledgments}

\end{document}